\newif\ifacm\acmfalse
\ifacm
\documentclass[letterpaper]{sig-alternate}
\else
\documentclass[11pt,letterpaper]{article}
\fi

\ifacm \else

\usepackage{amsthm}
\usepackage[margin=1in,dvips]{geometry}
\fi

\usepackage{graphicx}
\usepackage{amssymb}
\usepackage[boxed,section]{algorithm}
\usepackage{algpseudocode}
\usepackage{amsmath}
\usepackage{float}
\usepackage{enumerate}

\newtheorem{theorem}{Theorem}[section]
\newtheorem{lemma}[theorem]{Lemma}
\newtheorem{corollary}[theorem]{Corollary}
\newtheorem{definition}[theorem]{Definition}
\newtheorem{claim}[theorem]{Claim}

\newcommand{\abs}[1]{\left|#1\right|}
\newcommand{\floor}[1]{\left\lfloor#1\right\rfloor}
\newcommand{\norm}[2]{\left \lVert#2\right \rVert_{#1}}

\newcommand{\emd}{\text{EMD}}

\newcommand{\Mc}{ {\cal M}}
\newcommand{\Sc}{{\cal S}}
\newcommand{\R}{{\mathbb R}}
\newcommand{\Tc}{{\cal T}}

{\makeatletter
 \gdef\xxxmark{%
   \expandafter\ifx\csname @mpargs\endcsname\relax 
     \expandafter\ifx\csname @captype\endcsname\relax 
       \marginpar{xxx}
     \else
       xxx 
     \fi
   \else
     xxx 
   \fi}
 \gdef\xxx{\@ifnextchar[\xxx@lab\xxx@nolab}
 \long\gdef\xxx@lab[#1]#2{{\bf [\xxxmark #2 ---{\sc #1}]}}
 \long\gdef\xxx@nolab#1{{\bf [\xxxmark #1]}}
}

\DeclareMathOperator*{\argmax}{arg\,max}
\DeclareMathOperator*{\argmin}{arg\,min}
\DeclareMathOperator{\supp}{supp}
\DeclareMathOperator{\E}{E}

\def\R{\mathbb{R}}
\def\eps{\epsilon}

\begin{document}

\title{K-Median Clustering, Model-Based Compressive Sensing, and Sparse Recovery for Earth Mover Distance\thanks{This research has been supported in part by the David and Lucille
Packard Fellowship, MADALGO (the Center for Massive Data Algorithmics,
funded by the Danish National Research Association) and NSF grant
CCF-0728645.  E. Price has been supported in part by an NSF Graduate
Research Fellowship.
}}

\ifacm
\else

\author{Piotr Indyk\and Eric Price}
\date{24 April 2011}
\thispagestyle{empty}

\fi

\maketitle

\begin{abstract}
We initiate the study of sparse recovery problems under the {\em
  Earth-Mover Distance (EMD)}.  Specifically, we design a distribution
  over $m \times n$ matrices $A$ such that for any $x$, given $Ax$, we
  can recover a $k$-sparse approximation to $x$ under the EMD
  distance. One construction yields $m=O(k \log (n/k))$ and a $1
  + \eps$ approximation factor, which matches the best achievable
  bound for other error measures, such as the $\ell_1$ norm.

Our algorithms are obtained by exploiting novel connections to other
problems and areas, such as {\em streaming algorithms for k-median
  clustering} and {\em model-based compressive sensing.} We also provide novel algorithms and results  for the latter problems. 
\end{abstract} 

\ifacm
\else 
\fi

\section{Introduction}\label{sec:introduction}

In recent years, a new ``linear'' approach for obtaining a succinct
approximate representation of $n$-dimensional vectors (or signals) has
been discovered.  For any signal $x$, the representation is equal to
$Ax$, where $A$ is an $m \times n$ matrix, or possibly a random
variable chosen from some distribution over such matrices.  The vector
$Ax$ is often referred to as the {\em measurement vector} or {\em
  linear sketch} of $x$.  Although $m$ is typically much smaller than
$n$, the sketch $Ax$ often contains plenty of useful information about
the signal $x$.

A particularly useful and well-studied problem is that of {\em stable
  sparse recovery}.  The problem is typically defined as follows: for
some norm parameters $p$ and $q$ and an approximation factor $C>0$,
given $Ax$, recover an ``approximation'' vector $x^*$ such that
\begin{equation}
\label{e:lplq}
\norm{p}{x-x^*} \le C \min_{k\text{-sparse } x'}  \norm{q}{x-x'}
\end{equation}
where we say that $x'$ is $k$-sparse if it has at most $k$ non-zero
coordinates.  Sparse recovery has applications to numerous areas such
as data stream computing~\cite{Muthu:survey2, I-SSS} and compressed
sensing~\cite{CRT06:Stable-Signal, Don06:Compressed-Sensing}, notably
for constructing imaging systems that acquire images directly in
compressed form~(e.g.,~\cite{DDTLTKB,Rom}).  The problem has been a
subject of extensive study over the last few years, with the goal of
designing schemes that enjoy good ``compression rate'' (i.e., low
values of $m$) as well as good algorithmic properties (i.e., low
encoding and recovery times).  It is known by now\footnote{In
  particular, a random Gaussian matrix~\cite{CRT06:Stable-Signal} or a
  random sparse binary matrix~(\cite{BGIKS08}, building on
  \cite{CCF,CM03b,CM06}) has this property with overwhelming
  probability.  See~\cite{GI} for an overview.} that there exist
matrices $A$ and associated recovery algorithms that produce
approximations $x^*$ satisfying Equation~\eqref{e:lplq} with
$\ell_p=\ell_q=\ell_1$, constant approximation factor $C$, and sketch
length $m=O(k \log (n/k))$; it is also known that this sketch length
is asymptotically optimal~\cite{DIPW,FPRU}.  Results for other
combinations of $\ell_p$/$\ell_q$ norms are known as well.

However, limiting the error measures to variants of $\ell_p$ norms is
quite inconvenient in many applications. First, the distances induced
by $\ell_p$ norms are typically only quite raw approximations of the
perceptual differences between images. As a result, in the field of
computer vision, several more elaborate notions have been proposed
(e.g., in~\cite{RTG,Low,Lyu,GD}).  Second, there are natural classes
of images for which the distances induced by the $\ell_p$ norm are
virtually meaningless. For example, consider images of ``point
clouds'', e.g., obtained via astronomical imaging.  If we are given
two such images, where each point in the second image is obtained via
small random translation of a point in the first image, then the
$\ell_p$ distance between the images will be close to the largest
possible, even though the images are quite similar to each other.

Motivated by the above considerations, we initiate the study of sparse
recovery under non-$\ell_p$ distances. In particular, we focus on the
{\em Earth-Mover Distance} (EMD)~\cite{RTG}. Informally, for the case
of two-dimensional $\Delta \times \Delta$ images (say, $x,y :
[\Delta]^2 \to \R_+$) which have the same $\ell_1$ norm, the EMD is
defined as the cost of the min-cost flow that transforms $x$ into $y$,
where the cost of transporting a unit of mass from a pixel $p \in
[\Delta]^2 $ of $x$ to a pixel $q \in [\Delta]^2$ of $y$ is equal to
the $\ell_1$ distance\footnote{One can also use the $\ell_2$ distance.
  Note that the two distances differ by at most a factor of $\sqrt{2}$
  for two-dimensional images.} between $p$ and $q$.  The EMD metric
can be viewed as induced by a norm $\norm{EMD}{\cdot}$, such that
$\emd(x,y)=\norm{EMD}{x-y}$; see Section~\ref{sec:preliminaries} for a
formal definition. Earth-Mover Distance and its variants are popular
metrics for comparing similarity between images, feature sets,
etc.~\cite{RTG,GD}.

{\bf Results.} In this paper we introduce three sparse recovery
schemes for the Earth-Mover Distance.  Each scheme provides a matrix
(or a distribution of matrices) $A$, with $m$ rows and $n$ columns for
$n=\Delta^2$, such that for any vector $x$, given $Ax$, one can
reconstruct a vector $x^*$ such that
 \begin{equation}
\label{e:emd}
\norm{EMD}{x - x^*} \le C \min_{k\text{-sparse } x'} \norm{EMD}{x-x'}.
\end{equation}
 for some approximation factor $C > 0$.  We call any recovery scheme
 satisfying Equation~\eqref{e:emd} an \emph{EMD/EMD recovery scheme}.
 If $A$ is a distribution over matrices (that is, the scheme is {\em
   randomized}), the guarantee holds with some probability.  The other
 parameters of the constructions are depicted in
 Figure~\ref{f:results}.

\begin{figure}
\begin{center}
\begin{tabular}{|c|c|c|c|}
\hline
Determinism	& Sketch length &	Decode time & 	Approx. \\
 \hline
 Deterministic & $k \log n \log(n/k)$ & $n \log^{O(1)} n$ & $\epsilon$\\
 Deterministic & $k  \log(n/k)$ & $n^{O(1)}$ & $\sqrt{\log (n/k)}$\\
 Randomized  & $k \log(n/k)$ & $k \log(n/k)$ & $\epsilon$\\
 \hline
 \end{tabular}
 \caption{Our results}
 \label{f:results}
 \end{center}
 \end{figure}

In particular, two of our constructions yield sketch lengths $m$
bounded by $O(k \log (n/k) )$, which mimics
 the best possible bound
achievable for sparse recovery in the $\ell_1$ distance~\cite{DIPW}. Note, however, that we are not able to show a matching lower bound for the EMD case.

\paragraph{Connections and applications}  What does sparse recovery with
respect to the Earth-Mover Distance mean? Intuitively, a sparse
approximation under EMD yields a short ``signature'' of the image $x$
that approximately preserves its ``EMD properties''.  For example, if
$x$ consists of a small number of sparse point clouds (e.g., as in
astronomical imaging), sparse approximation of $x$ will approximately
identify the locations and weights of the clouds.  Our preliminary
experiments with a heuristic algorithm for such data~\cite{GIP} show
that this approach can yield substantial improvements over the usual
sparse recovery. 

Another application~\cite{RTG} stems from the
original paper, where such short signatures were
constructed\footnote{In fact, the algorithm in~\cite{RTG} vaguely
  resembles our approach, in that it uses a kd-tree decomposition to
  partition the images.} for general images, to extract their color or
texture information. The images were then replaced by their signatures
during the experiments, which significantly reduced the computation
time.

The above intuitions can be formalized as follows.
 Let $x'$ be the minimizer of $
\norm{EMD}{x - x'}$ over all $k$-sparse vectors. Then one can observe that the
non-zero entries of $x'$ correspond to the cluster centers in the best
$k$-median\footnote{For completeness, in our context the $k$-median is
  defined as follows. First, each pixel $p \in [\Delta]^2$ is interpreted
  as a point with weight $x_p$.
  The goal is to find a set $C \subset [n]^2$ of $k$ ``medians'' that
  minimizes the objective function $\sum_{p \in [n]^2} \min_{c \in C}
  \norm{1}{p-c} x_p$.} clustering of $x$.  Moreover, for each such center
$c$, the value of $x'_c$ is equal to the total weight of pixels in the
cluster centered at $c$.  Thus, a solution to the $k$-median problem provides a solution to our sparse recovery
problem as well\footnote{If the algorithm reports both the medians and the weights of clusters.}.

There has been prior work on the $k$-median problem in the streaming model
under insertions and deletions of points~\cite{FSoh,I04}. Such
algorithms utilize linear sketches, and therefore implicitly provide
schemes for approximating the $k$-median of $x$ from a linear sketch of
$x$ (although they do not necessarily provide the cluster weights,
which are needed for the sparse recovery problem).  Both
algorithms\footnote{The paper~\cite{I04} claims $m = k \log^{O(1)} n$.
  Unfortunately, that is an error, caused by ignoring the dependencies
  between the queries and their answers provided by the randomized
  data structure {\sc MediEval}. Fixing this problem requires reducing
  the probability of failure of the algorithm so that it is inversely
  exponential in $k$, which yields another factor of $k$ in the space
  bound.} yield a method for approximating the $k$-median from
$\Omega(k^2\log^{O(1)} n)$ measurements, with the algorithm of~\cite{FSoh}
providing an approximation factor of $1+\epsilon$. In contrast, our
result achieves an approximation factor of $1+\epsilon$ with a sketch
length $m$ that is as low as $O(k \log(n/k) )$.

Thanks to this connection, our results also yield short sketches for
the $k$-median problem.  Although the solution $x^*$ output by our
algorithm does not have to be $k$-sparse (i.e., we might output more
than $k$ medians), one can post-process the output by computing the
best $k$-sparse approximation to $x^*$ using any off-the-shelf
(weighted) $k$-median algorithm (e.g.,~\cite{HM})).  This reduces the
number of clusters to $k$, while (by the triangle inequality of EMD) multiplying the approximation factor
by a constant that depends on the approximation constant of the chosen
$k$-median algorithm. See Appendix~\ref{s:strict} for more details.



\paragraph{Techniques} On a high level, our approach is to reduce the
sparse recovery problem under EMD to sparse recovery under
$\ell_1$. This is done by constructing a linear mapping $P$ that maps
$\R^{[\Delta]^2}$ into some space $\R^t$, that has the property that a
``good'' sparse approximation to $y=Px$ under $\ell_1$ yields a
``good'' sparse approximation to $x$ under EMD. \footnote{We note that the aforementioned k-median algorithms implicitly rely on some form
 of sparse recovery (e.g., see Remark 3.10 in~\cite{FSoh} or remarks
 before Theorem 5 in~\cite{I04}).  However, the bounds provided by
 those algorithms fall short of what we aim for.}  The list of formal
constraints that such a mapping needs to satisfy are given in
Section~\ref{sec:framework}.  For concreteness, we define one such
mapping below; another one is given in Section~\ref{sec:haar}.
Specifically, the \emph{pyramid} mapping $P$ ~\cite{IT,GD} (building
on~\cite{Char,AV99}) is defined as follows.  First we impose $\log
\Delta+1$ nested grids $G_i$ on $[\Delta]^2$, with $G = \bigcup G_i$.
For each \emph{level} $i=0 \ldots l$, $l=\log_2 \Delta$, the grid $G_i$ is
a partition of the image into \emph{cells} of side length $2^i$.  The
cells in the grids can be thought of as forming a $4$-ary tree, with
each node $c$ at level $i$ having a set $C(c)$ of children at level
$i-1$.  For each $i$, we define a mapping $P_i$ such that each entry
in $P_i x$ corresponds to a cell $c$ in $G_i$, and its value is equal
to the sum of coordinates of $x$ falling into $c$.  The final mapping
$P$ is defined as
\begin{eqnarray}
\label{e:pyramid}
 Px = [ 2^0 P_0 x, 2^1 P_1x, \ldots, 2^l P_l x] 
 \end{eqnarray}

It is easy to see that, for a vector $x$ that is $k$-sparse, the
vector $Px$ is $O(K)$ sparse for $K=kl$.  We also show that for {\em
  any} $x$, there exists an $O(K)$-sparse $y$ such that the difference
$\|y-Px\|_1$ is comparable to $\min_{k\text{-sparse } x'}
\norm{EMD}{x-x'}$.  We then find a good approximation $x^*$ to $x$
(in the EMD norm) by ``inverting'' $P$ on $y$.  Since we can recover
an $O(K)$-sparse approximation to $y$ (in the $\ell_1$ norm) from a
sketch of length $O(K\log(n/K))$, we obtain the first result from
Figure~\ref{f:results}.

To improve the sketch length we exploit the particular properties of
the mapping $P$ to recover an $O(K)$-sparse approximation from only
$O(K)$ measurements.  For any non-negative vector $x$, the coordinates
of $Px$ have the following hierarchical structure: (i) the coordinates
are organized into an $r$-ary tree for $r=4$, and (ii) the value of
each internal node is non-negative and equal to the sum of its
children {\em times two}. Using one or both of these properties
enables us to reduce the number of measurements.

 The second algorithm from Figure~\ref{f:results} is obtained using
 the property (i) alone.  Specifically, the problem of recovering a
 sparse approximation whose support forms a tree has been well-studied
 in signal processing (the question is motivated by an empirical
 observation that large wavelet coefficients tend to co-occur in this
 fashion).  In particular, the  insightful paper~\cite{BCDH} on {\em
   model-based compressive sensing} (see Section~\ref{sec:model-based}
 for an overview) gave a deterministic scheme that recovers such
 approximation from a sketch of length $O(K)$. Although the setup
 given in that paper is somewhat different from what we need here, we
 show that one can modify and re-analyze their scheme to achieve the
 desired guarantee.  This approach, however, leads to an approximation
 factor of $O(\sqrt{\log(n/k)})$.
 
 In order to achieve a constant approximation factor, we employ both
 properties (i) and (ii), as well as randomization.  Specifically, we
 recover the tree coefficients top-down, starting from the root of the
 tree.  This is done in a greedy manner: we only recurse on the children
 of nodes that are estimated to be ``heavy''.  This first pass
 identifies a superset $S$ of the locations where $Px$ is large, but
 estimates some of the values $(Px)_S$ quite poorly.  The set of locations $S$
 has $\abs{S} = O(K)$, so we can recover $(Px)_S$ accurately with
 $O(K)$ measurements using the {\em set query sketches}
 of~\cite{Pri11}.
 
Finally, we show that we can achieve the first and second result in
Figure~\ref{f:results} by replacing the pyramid mapping by a variant
of an even more basic transform, namely the
(two-dimensional) Haar wavelet mapping.  Our variant is obtained by
rescaling the original Haar wavelet vectors using exponential weights,
to mimic the pyramid scheme behavior.  This result relates the two
well-studied notions (EMD and wavelets) in a somewhat unexpected way.
As a bonus, it also simplifies the algorithms, since inverting the
wavelet mapping can now be done explicitly and losslessly.

 \floatstyle{boxed} 
\restylefloat{figure}

 \section{Preliminaries}\label{sec:preliminaries}

\paragraph{Notation}
We use $[n]$ to denote the set $\{1 \ldots n\}$.  For any set $S
\subset [n]$, we use $\overline{S}$ to denote the complement of $S$,
i.e., the set $[n]\setminus S$.  For any $x \in \R^n$, $x_i$ denotes
the $i$th coordinate of $x$, and $x_S$ denotes the vector $x' \in
\R^n$ given by $x'_i = x_i$ if $i \in S$, and $x'_i = 0$ otherwise.
We use $\supp(x)$ to denote the support of $x$.  We use
$\R^{[\Delta]^2}$ to denote the set of functions from $[\Delta] \times
[\Delta]$ to $\R$; note that $\R^{[\Delta]^2}$ can be identified with
$\R^n$ since $n=\Delta^2$.  We also use $\R_+$ to denote $\{x \in \R
\mid x \geq 0\}$.

\paragraph{EMD} Consider any two
non-negative vectors $x,y \in \R_+^{[\Delta]^2}$ such that
$\|x\|_1=\|y\|_1$.  Let $\Gamma(x,y)$ be the set of functions $\gamma:
[\Delta]^2 \times [\Delta]^2 \to \R_+$, such that for any $i,j \in
[\Delta]^2$ we have $\sum_l \gamma(i,l)=x_i$ and $\sum_l
\gamma(l,j)=y_{j}$; that is, $\Gamma$ is the set of possible ``flows''
from $x$ to $y$.  Then we define
\[ \emd^*(x,y) = \min_{\gamma \in \Gamma} \sum_{i,j \in [\Delta]^2 } \gamma(i,j)\|i-j\|_1 \]
to be the min cost flow from $x$ to $y$, where the cost of an edge is
its $\ell_1$ distance.  This induces a norm $\norm{EMD}{\cdot}$ such
that $\norm{EMD}{x-y} = \emd^*(x, y)$.  For general vectors $w$,
\[
\norm{EMD}{w} = \min_{\substack{x - y + z = w\\\norm{1}{x}=\norm{1}{y}\\x, y \geq 0}} \emd^*(x, y) + D\norm{1}{z}
\]
where $D=2\Delta$ is the diameter of the set $[\Delta]^2$.  That is,
$\norm{EMD}{w}$ is the min cost flow from the positive coordinates of
$w$ to the negative coordinates, with some penalty for unmatched mass.

\paragraph{Signal models}
The basic idea of the signal models framework of~\cite{BCDH} is to
restrict the sparsity patterns of the approximations.  For some
sparsity parameter\footnote{We use $K$ to denote the sparsity in the
  context of model-based recovery (as opposed to $k$, which is used in
  the context of ``standard'' recovery).} $K$ let $\Sc_K$ be a family
of subsets of $[n]$ such that for each $S \in \Sc_K$ we have $|S| \le
K$.  The family $\Sc_K$ induces a {\em signal model} $\Mc_K \subset
\R^n$ where
\[ \Mc_K =  \{ x \in \R^n\mid \supp(x) \subseteq S \text{\ for some\ } S \in \Sc_K \}. \]
Note that $\Mc_K$ is a union of $|\Sc_K|$ subspaces, each of dimension
at most $K$.  The signals in $\Mc_K$ are called $\Mc_K$-{\em sparse}.

The following two examples of signal models are particularly relevant
to our paper:
\begin{enumerate}
\item General $k$-sparse signals, where $\Sc_k$ contains all $k$-subsets
  of $[n]$. In this case the induced signal model (denoted by
  $\Sigma_k$) contains all $k$-sparse signals.
\item Tree sparse signals. In this case, we assume that $n=
  \frac{c^l-1}{c-1}$ for some (constant) integer $c$ and parameter
  $l$, and associate each $i \in [n]$ with a node of a full $c$-ary
  tree $T(c,l)$ of depth $l$.  The family $\Sc_K$ contains all sets
  $S$ of size up to $K$ that are connected in $T(c,l)$ and contain the
  root (so each $S$ corresponds to a graph-theoretic subtree of $T(c,
  l)$).  The induced signal model is denoted by $\Tc^c_K$, or $\Tc_K$
  for short.\footnote{We note that technically this model was
    originally defined with respect to the wavelet basis (as opposed
    to the standard basis here) and for $c=2$. We adapt that
    definition to the needs in our paper.}
\end{enumerate}

In order to facilitate signal recovery, one often needs to consider
the differences $x-y$ of two signals $x \in \Mc$, $y \in \Mc'$.  For
this purpose we define the Minkowski sum of $\Mc_K$ and $\Mc'_K$ as
$\Mc_K \oplus \Mc'_K= \{x+y: x \in \Mc_K, y \in \Mc'_K\}$.  To
simplify the notation, we define $\Mc^{(t)} $ to the $t$-wise
Minkowski sum of $\Mc_K$.  For all signal models considered in this
paper, we have $\Mc_K^{(t)} \subset \Mc_{Kt}$.

Restricting sparsity patterns enables to recover sparse approximations
from shorter sketches. We defer a more thorough overview of the
results to Section~\ref{sec:model-based}.

\paragraph{Assumptions}
We assume that the sparsity parameters $k$ (and $K$, where applicable)
are smaller than $n/2$.  Note that if this assumption does not hold,
the problem becomes trivial, since one can define the measurement
matrix $A$ to be equal to the identity matrix.

\section{Framework for EMD-sparse recovery}\label{sec:framework}

 In this section we describe our approach to reducing sparse recovery
 under $\emd$ into sparse recovery under $\ell_1$.  We need the
 following three components: (i) a $t \times n$ matrix $B$ (that will
 be used to map the $\emd$ space into the $\ell_1$ space); (ii) a
 signal model $\mathcal{M} \subset \R^t$; and (iii) an $\ell_1/\ell_1$
 recovery scheme for $\Mc$. The latter involves an $m \times t$ matrix
 $A'$ (or a distribution over such matrices) such that, for any $x \in
 \R^t$, given $A'x$, one can recover $x^*$ such that
 \begin{equation}
\label{e:model}
\norm{1}{x-x^*} \le C' \min_{x' \in \Mc} \norm{1}{x-x'}
\end{equation}
for an approximation factor $C'$.  If $A'$ is a distribution over
matrices, we require that the guarantee holds with some constant
probability, e.g., 2/3.

\bigskip\noindent
  The mapping $B$ must satisfy the following three properties:
  \begin{enumerate}[A.]
    \item\label{prop:expansion} (EMD-to-$\ell_1$ expansion.) For all $x \in
      \R^n$, $$\norm{EMD}{x} \leq \norm{1}{Bx}.$$

    \item\label{prop:model-alignment} (Model-alignment of EMD with $\Mc$.)  For
      all $x \in \R_+^{n}$, there exists a $y \in \mathcal{M}$ with
    \[
    \norm{1}{y-Bx} \leq \eps \min_{k\text{-sparse } x'}\norm{EMD}{x - x'}.
    \]

    \item\label{prop:invertibility} (Invertibility.) There is an
      efficient algorithm $\mathcal{B}^{-1} \colon \R^t \to \R^{n}$ such
      that, for some constant $D$ and all $y \in \R^t$,
      \[
      \norm{1}{y-B\mathcal{B}^{-1}(y)} \leq D \min_{x \in \R^{n}} \norm{1}{y-Bx}.
      \]

  \end{enumerate}
  \bigskip

\begin{lemma}\label{lemma:framework}
  Consider $B,A',\Mc$ satisfying the above properties.
  Then the matrix $A=A' B$ supports $k$-sparse recovery for EMD (as defined in Equation~\eqref{e:emd}) with approximation factor $C=(1+D)C'\eps$.
\end{lemma}
\begin{proof}
  Consider the recovery of any vector $x \in \R_+^{n}$.  Let
  \[
  E = \min_{k\text{-sparse } x'}\norm{EMD}{x - x'}.
  \]
  By Property~\ref{prop:model-alignment}, for any $x \in \R^n$, there
  exists a $y \in \mathcal{M}$ with
  \[
  \norm{1}{y-Bx} \leq \eps E.
  \]
  Hence our $\ell_1/\ell_1$ model-based recovery scheme for $\Mc$,
  when run on $Ax = A'Bx$, returns a $y^*$ with
  \[
  \norm{1}{y^*-Bx} \leq C' \eps E.
  \]
  Let $x^* = \mathcal{B}^{-1}(y^*)$.  We have by
  Property~\ref{prop:invertibility} that
  \[
  \norm{1}{y^*-Bx^*} \leq D \min_{x' \in \R^{n}} \norm{1}{y^*-Bx'}
  \leq D\norm{1}{y^*-Bx} \leq D C' \eps E.
  \]
  Hence by Property~\ref{prop:expansion}
  \begin{align*}
    \norm{EMD}{x^*-x} &\leq \norm{1}{B(x^*-x)} \leq \norm{1}{y^*-Bx} +
    \norm{1}{y^*-Bx^*} \\& \leq (1+D)C' \eps E
  \end{align*}
  as desired.
\end{proof}

\section{Pyramid transform}

In this section we will show that the pyramid transform $P$ defined in
Equation~\eqref{e:pyramid} of Section~\ref{sec:introduction} satisfies
properties~\ref{prop:model-alignment} and~\ref{prop:invertibility}
of Section~\ref{sec:framework}, with appropriate parameters.

The property~\ref{prop:expansion} has been shown to hold for $P$ in many other papers
(e.g.,~\cite{Char,IT}).  The intuition is that the weight of a cell is
at least the Earth-Mover Distance to move all mass in the cell from
the center to any corner of the cell, including the corner that is at
the center of the parent of the cell.

\subsection{Model-alignment with tree sparsity}

In this section we show Property~\ref{prop:model-alignment}, where the signal model $\Mc$ is
equal to the $K$-tree-sparse model $\Tc_K$, for $K=O(k \log(n/k))$.
In fact, we show a stronger statement: the trees have their
\emph{width} (the maximum number of nodes per level) bounded by some
parameter $s$.  We will exploit the latter property later in the
paper.

\begin{lemma}\label{lemma:pyramid-model}\label{goodtopdown}
 For any $x \in \R_+^n$  there exists  a tree $S \subset [t]$ of size $K$ and width $s$ with
  \[
  \norm{1}{(Px)_{\overline{S}}} \leq \eps \min_{k\text{-sparse }x'} \norm{EMD}{x - x'}
  \]
  for $s = O\left(\frac{1}{\eps^2}k\right)$ and $K =
  O(\frac{1}{\eps^2}k\log (n/k))$.
\end{lemma}
\begin{proof}
  Let $x' = \argmin_{k\text{-sparse } x'} \norm{EMD}{x-x'}$ be the
  $k$-medians approximation of $x$.  Consider the cells that contain
  each point in the support of $x'$.  For each such cell at any level
  $i$, add the $O(\frac{1}{\eps^2})$ other cells of the same level
  within an $\ell_1$ distance of $\frac{2}{\eps}2^i$.  The resulting
  $S$ has $s=O\left(\frac{1}{\eps^2}k\right)$ cells per level, and all
  the ancestors of any cell in the result also lie in $S$.  So $S$ is
  a tree of width $s$.  It has $O(s)$ elements from the top $\log_4 s$
  levels, and $O(s)$ elements on each of the $\log_4 t - \log_4 s$
  remaining levels, for a size $K = O(s \log t/s)$.  We will show that
  $\norm{1}{(Px)_{\overline{S}}}$ is small.

  Define $e_i$ for $i \in [\Delta]^2$ to be the elementary vector with
  a $1$ at position $i$, so $x_i = x \cdot e_i$.  Suppose that
  the distance between $i$ and the nearest center in $x'$ is $v_i$.
  Then we have
  \begin{align*}
    \norm{1}{(Px)_{\overline{S}}} &= \sum_{i\in[\Delta]^2} \norm{1}{(Px_ie_i)_{\overline{S}}} = \sum_{i\in[\Delta]^2} \norm{1}{(Pe_i)_{\overline{S}}}x_i\\
    \norm{EMD}{x - x'} &= \sum_{i\in[\Delta]^2} v_ix_i.
  \end{align*}
  so it is sufficient to show $\norm{1}{(Pe_i)_{\overline{S}}} \leq \eps
  v_i$ for any $i$.

  Let $h$ be the highest level such that $e_i$ is not contained in a
  cell at level $h$ in $S$.  If no such $h$ exists,
  $\norm{1}{(Pe_i)_{\overline{S}}} = 0$.  Otherwise, $v_i \geq
  \frac{2}{\eps}2^{h}$, or else $S$ would contain $e_i$'s cell in
  level $h$.  But then
  \begin{align*}
    \norm{1}{(Pe_i)_{\overline{S}}} = \sum_{j=0}^{h} 2^j &< 2^{h+1} \leq \eps v_i
  \end{align*}
  as desired.
\end{proof}
\begin{corollary}
  For any $x \in \R_+^n$, there exists a $y \in \mathcal{T}_K$ with
  \[
    \norm{1}{y-Px} \leq \eps \min_{k\text{-sparse } x'}\norm{EMD}{x - x'}.
  \]
\end{corollary}

\subsection{Invertibility}\label{sec:invertibility}


Given an approximation $b$ to $Px$, we would like to find a vector $y$
with $\norm{1}{b-Py}$ small.  Note that this task can be formulated as
a linear program, and therefore solved in time that is polynomial in
$n$.  In Appendix~\ref{app:invertibility} we show a much faster {\em
  approximate} algorithm for this problem, needed for our fast
recovery algorithm:

\begin{lemma}\label{lemma:pyramidinverse}
  Given any approximation $b$ to $Px$, we can recover a $y$ in
  $O(\abs{\supp(b)})$ time with
  \[
  \norm{1}{Py-Px} \leq 8\norm{1}{b-Px}.
  \]
\end{lemma}

Recall that $P$ has $t = \floor{4n/3}$ rows.  This means standard
$\ell_1/\ell_1$ $K$-sparse recovery for $Px$ is possible with $m = O(K
\log t/K) = O(\frac{1}{\eps^2}k \log^2 (n/k))$.  Hence by
Lemma~\ref{lemma:framework}, using $B=P$ and standard sparse recovery
techniques on the model $\Mc = \Sigma_K$ gives the first result in
Figure~\ref{f:results}:
\begin{theorem}
  There exists a deterministic EMD/EMD recovery scheme with $m =
  O(\frac{1}{\eps^2}k \log^2(n/k))$ and $C = \eps$.
  Recovery takes $O(n \log^c n)$ time for some constant $c$.
\end{theorem}

\section{Tree-sparse recovery}\label{sec:model-based}

To decrease the number of measurements required by our algorithm, we
can use the stronger signal model $\mathcal{T}_K$ instead of
$\Sigma_K$.  The paper~\cite{BCDH} gives an algorithm for model-based
sparse recovery of $\mathcal{T}_K$, but their theorem does not give an
$\ell_1/\ell_1$ guarantee.  In Appendix~\ref{app:model-based} we
review the prior work and convert their theorem into the following:

\begin{theorem}
There exists a matrix $A$ with $O(K)$ rows and a recovery algorithm
that, given $Ax$, returns $x^*$ with
\[
\norm{1}{x-x^*} \le C \sqrt{\log(n/K)}  \min_{x' \in \Tc_K} \norm{1}{x-x'}
\]
for some absolute constant $C>1$. As long as the coefficients of $x$
are integers bounded by $n^{O(1)}$, the algorithm runs in time $O(K^2n
\log^{c} n)$ for some constant $c$.
\label{t:l1l1model}
\end{theorem}
By Lemma~\ref{lemma:framework}, using this on $B=P$ and $\Mc =
\mathcal{T}_K$ gives the second result in Figure~\ref{f:results}:
\begin{theorem}
  There exists a deterministic EMD/EMD recovery scheme with $m =
  O(\frac{1}{\eps^2}k \log(n/k))$ and distortion $C =
  O(\eps\sqrt{\log(n/k)})$.  Recovery takes $O(k^2n \log^c n)$ time
  for some constant $c$.
\end{theorem}

\section{Beyond tree sparsity}\label{sec:randomized}

The previous section achieved $O(\sqrt{\log n})$ distortion
deterministically with $O(k \log (n/k))$ rows.  In this section, we
improve the distortion to an arbitrarily small constant $\eps$ at the
cost of making the algorithm randomized.  To do this, we show that EMD
under the pyramid transform is aligned with a stronger model than just
tree sparsity---the model can restrict the values of the coefficients
as well as the sparsity pattern.  We then give a randomized algorithm
for $\ell_1/\ell_1$ recovery in this model with constant distortion.

\begin{definition}
  Define $T_K^s$ to be the family of sets $S \subseteq [t]$ such that (i)
  $S$ corresponds to a connected subset of $G$ containing the root and
  (ii) $\abs{S \cap G_i} \leq s$ for all $i$.  We say that such an $S$
  is \emph{$K$-tree-sparse with width $s$}.
\end{definition}
\begin{definition}
  Define $\mathcal{M} \subset \mathcal{T}_K$ as
  \[
  \mathcal{M} = \left\{y \in \R^t \middle| \begin{array}{l}\supp(y) \subseteq S\text{ for some $S \in T_K^s$, and}\\y_i \geq 2\norm{1}{y_{C(i)}} \forall i \in [t]\end{array}\right\}.
  \]
  where $s = O(\frac{1}{\eps^2}k)$ comes from
  Lemma~\ref{lemma:pyramid-model}.
\end{definition}
Note that every $y \in \mathcal{M}$ is non-negative, and $(Px)_S \in
\mathcal{M}$ for all $x \in \R_+^n$.  With
Lemma~\ref{lemma:pyramid-model}, this implies:

\begin{lemma}
  There is model-alignment of $P$ with $\Mc$, i.e., they satisfy
  Property~\ref{prop:model-alignment}.
\end{lemma}

We will give a good algorithm for $\ell_1/\ell_1$ recovery over
$\mathcal{M}$.

\subsection{Randomized $\ell_1/\ell_1$ recovery of $\mathcal{M}$}

\begin{theorem}\label{thm:randomizedl1l1}
  There is a randomized distribution over $m \times t$ matrices $A$
  with $m = O(\frac{1}{\eps^2}k\log (n/k))$ and an algorithm that
  recovers $y^*$ from $Ay$ in $O(\frac{1}{\eps^2}k\log (n/k))$ time
  with
  \[
  \norm{1}{y^* - y} \leq C \min_{y' \in \mathcal{M}} \norm{1}{y - y'}
  \]
  with probability $1-k^{-\Omega(1)}$, for some constant $C$.  We assume $k
  = \Omega(\log \log n)$.
\end{theorem}

We will give an algorithm to estimate the support of $y$.  Given a
sketch of $y$, it recovers a support $S \in T_K^{2s}$ with
\[
\norm{1}{y_{\overline{S}}} \leq 10 \min_{y' \in \mathcal{M}}\norm{1}{y - y'}.
\]
We can then use the set query algorithm~\cite{Pri11} to recover a
$y^*$ from a sketch of size $O(\abs{S})$ with
\[
\norm{1}{y^* - y_S} \leq \norm{1}{y_{\overline{S}}}.
\]
Then
\[
\norm{1}{y^*-y} \leq \norm{1}{y^*-y_S}+\norm{1}{y-y_S}\leq 2\norm{1}{y_{\overline{S}}} \leq 20\min_{y' \in \mathcal{M}}\norm{1}{y - y'}.
\]
as desired.  Hence estimating the support of $y$ is sufficient.

\subsection{Finding a good sparse support $S$ to $y$}

Vectors $y' \in \mathcal{M}$ have two properties that allow us to find
good supports $S \in T_K^s$ with constant distortion using only
$O(\abs{S})$ rows.  First, $\supp(y')$ forms a tree, so the support
can be estimated from the top down, level by level.  Second, each
coefficient has value at least twice the sum of the values of its
children.  This means that the cost of making a mistake in estimating
the support (and hence losing the entire subtree below the missing
coefficient) is bounded by twice the weight of the missing
coefficient.  As a result, we can bound the global error in terms of
the local errors made at each level.

Of course, $y$ may not be in $\mathcal{M}$.  But $y$ is ``close'' to
some $y' \in \mathcal{M}$, so if our algorithm is ``robust'', it can
recover a good support for $y$ as well.  Our algorithm is described in
Algorithm~\ref{alg:findsupport}.

\begin{algorithm}
  \caption{Finding sparse support under
    $\mathcal{M}$}\label{alg:findsupport}

  \textbf{Definition of sketch matrix $A$.}  The algorithm is
  parameterized by a width $s$.  Let $h_i$ be a random hash function
  from $G_i$ to $O(s)$ for $i \in [\log(n/s)]$.  Then define $A'(i)$
  to be the $O(s)\times \abs{G_i}$ matrix representing $h_i$, so
  $A'(i)_{ab} = 1$ if $h_i(b) = a$ and $0$ otherwise.  Choose $A$ to
  be the vertical concatenation of the $A'(i)$'s.

  \textbf{Recovery procedure.}
  \begin{algorithmic}
    \\\Comment{Find approximate support $S$ to $y$ from $b = Ay$}
    \Procedure{FindSupport}{$A$, $b$}
    \State $T_{\log(n/s)} \gets G_{\log(n/s)}$
    \Comment{$\abs{T_{\log(n/s)}} \leq 2s$}
    \For{$i=\log(n/s)-1 \dotsc 0$}
    \\\Comment{Estimate $y$ over $C(T_{i+1})$.}
    \State $y^*_j \gets b_{h_i(j)}$ for $j \in C(T_{i+1})$.
    \\\Comment{Select the $2s$ largest elements of our estimate.}
    \State $T_i \gets {\displaystyle\argmax_{\substack{T' \subseteq C(T_{i+1})\\\abs{T'}\leq 2s}}} \norm{1}{y^*_{T'}}$
    \EndFor
    \State $S \gets \displaystyle\bigcup_{i=0}^{\log(n/s)} T_i \cup \bigcup_{i \geq \log(n/s)} G_i$
    \EndProcedure
  \end{algorithmic}

\end{algorithm}

\begin{lemma}\label{algksupporttime}
Algorithm~\ref{alg:findsupport} uses a binary sketching matrix
of $O(s\log(n/s))$ rows and takes $O(s\log(n/s))$ time to recover $S$
from the sketch.
\end{lemma}
\begin{proof}
  The algorithm looks at $O(\log (n/s))$ levels.  At each level it
  finds the top $2s$ of $4 \times 2s$ values, which can be done in
  linear time.  The algorithm requires a sketch with $O(\log (n/s))$
  levels of $O(s)$ cells each.
\end{proof}

The algorithm estimates the value of $y_{C(T_{i+1})}$ by hashing all
of $y_{G_i}$ into an $O(s)$ size hash table, then estimating $y_j$ as
the value in the corresponding hash table cell.  Since $y$ is
non-negative, this is an overestimate.  We would like to claim that
the $2s$ largest values in our estimate approximately contain the $s$
largest values in $y_{C(T_{i+1})}$.  In particular, we show that any
$y_j$ we miss is either (i) not much larger than $s$ of the
coordinates we do output or (ii) very small relative to the
coordinates we already missed at a previous level.

\begin{lemma}\label{lemma:algksupport}
  In Algorithm~\ref{alg:findsupport}, for every level $i$ let $w_i =
  \max_{q \in C(T_{i+1}) \setminus T_i} y_q$ denote the maximum value
  that is skipped by the algorithm and let $f_i = \norm{1}{y_{G_{i+1}
      \setminus T_{i+1}}}$ denote the error from coordinates not
  included in $T_{i+1}$.  Let $c_i$ denote the $s$-th largest value in
  $y_{T_i}$.  Then with probability at least $1 - e^{-\Omega(s)}$,
  $w_i \leq \max\{\frac{f_i}{4s}, 2c_i\}$ for all levels $i$.
\end{lemma}
\begin{proof}
  Define $s' = 8s \geq \abs{C(T_{i+1})}$.  We make the hash table size
  at each level equal to $u=32s'$.  We will show that, with high
  probability, there are at most $s$ coordinates $p$ where $y^*_p$ is more
  than $f_i/s'$ larger than $y_p$.  Once this is true, the result comes
  as follows: $y^*$ is an overestimate, so the top $2s$ elements of
  $y^*$ contain at least $s$ values that have been overestimated by at
  most $f_i/s'$.  Because the algorithm passes over an element of
  value $w_i$, each of these $s$ values must actually have value at
  least $w_i - f_i/s'$.  Hence either $w_i < 2f_i/s' = \frac{f_i}{4s}$
  or all $s$ values are at least $w_i/2$.

  To bound the number of badly overestimated coordinates, we split the
  noise in two components: the part from $G_i \setminus
  C(T_{i+1})$ and the part from $C(T_{i+1})$.  We will show that, with
  probability $1 - e^{-\Omega(s)}$, the former is at most $f_i/s'$ in
  all but $s/4$ locations and the latter is zero in all but $3s/4$
  locations.

  \textsc{WLOG} we assume that the function $h_i$ is first fixed for
  $G_i \setminus C(T_{i+1})$, then randomly chosen for $C(T_{i+1})$.
  Let $O_i \subset [u]$ be the set of ``overflow buckets'' $l$ such
  that the sum $s_l = \sum_{p \notin C(T_{i+1}), h_i(p)=l} y_p$ is at
  least $f_i/s'$.  By the definition of $f_i$, $\sum_l s_l = f_i/2$,
  so 
  \[ |O_i| / u \le \frac{f_i/2}{f_i/s'}/u = 1/2 \frac{s'}{32s'}=1/64 .\]
  Thus, the probability that a fixed child
  $q \in C(T_{i+1})$ is mapped to $O_i$ is at most $1/64$.  This is
  independent over $C(T_{i+1})$, so the Chernoff bound applies.  Hence
  with probability at least $1 - e^{-\Omega(s)}$, the number of $q\in
  C(T_{i+1})$ mapping to $O_i$ is at most twice its expectation, or
  $\abs{C(T_{i+1})}/32 = s/4$.

  We now bound the collisions within $C(T_{i+1})$.  Note that our
  process falls into the ``balls into bins'' framework, but for
  completeness we will analyze it from first principles.

  Let $Z$ be the number of cells in $C(T_{i+1})$ that collide.  $Z$ is
  a function of the independent random variables $h_i(p)$ for $p \in
  C(T_{i+1})$, and $Z$ changes by at most $2$ if a single $h_i(p)$
  changes (because $p$ can cause at most one otherwise non-colliding
  element to collide).  Hence by McDiarmid's inequality,
  \[
  \Pr[Z \geq \E[Z] + t] \leq e^{-t^2/(2s')}
  \]
  But we know that the chance that a specific $p$ collides with any of
  the others is at most  
  $s'/u = 1/32$.  Hence $\E[Z]
  \leq s'/32$, and
  \[
  \Pr[Z \geq (\frac{1}{32} + \eps)s'] \leq e^{-\eps^2s'/2}.
  \]
  By setting $\eps=2/32$ we obtain that,  with probability $1-e^{-\Omega(s)}$ we have that $Z \leq
  \frac{3s'}{32} = 3s/4$.

  Hence with probability $1 - e^{-\Omega(s)}$, only $3s/4$ locations
  have non-zero corruption from $C(T_{i+1})$, and we previously showed
  that with the same probability only $s/4$ locations are corrupted by
  $f'/s'$ from outside $C(T_{i+1})$.  By the union bound, this is true
  for all levels with probability at least $1 - (\log n)e^{-\Omega(s)}
  = 1 - e^{-\Omega(s)}.$
\end{proof}

\begin{lemma}\label{lemma:findsupport}
  Let $S$ be the result of running
  Algorithm~\ref{alg:findsupport} on $y \in \R^t$.  Then
  \[
  \norm{1}{y_{\overline{S}}} \leq 10 \min_{y' \in \mathcal{M}}\norm{1}{y - y'}
  \]
  with probability at least $1-e^{\Omega(s)}$.
\end{lemma}
\begin{proof}
  From the algorithm definition, $T_i = S \cap G_i$ for each level
  $i$.  Let $y' \in \mathcal{M}$ minimize $\norm{1}{y - y'}$, and let
  $U = \supp(y')$.  By the definition of $\mathcal{M}$, $U \in T_K^s$.

  For each $i$, define $V_i = U \cap C(T_{i+1}) \setminus T_{i}$ to be
  the set of nodes in $U$ that could have been chosen by the algorithm
  at level $i$ but were not.  For $q \in U \setminus S$, define $R(q)$
  to be the highest ancestor of $q$ that does not lie in $S$; hence
  $R(q)$ lies in $V_i$ for some level $i$.  Then
  \begin{align}
    \norm{1}{y'_{\overline{S}}} = \norm{1}{y'_{U\setminus S}} &= \sum_{q \in U \setminus S} y'_q\notag\\
    &= \sum_{i} \sum_{p \in V_i} \sum_{R(q) = p} y'_q\notag\\
    &\leq \sum_{i} \sum_{p \in V_i} 2y'_p\notag\\
    &= 2\sum_i \norm{1}{y'_{V_i}}, \label{eq:subtreesums}
  \end{align}
  where the inequality holds because each element of $y'$ is at
  least twice the sum of its children.  Hence the sum of $y'$
  over a subtree is at most twice the value of the root of the
  subtree.

  Define the error term $f_i = \norm{1}{y_{G_{i+1} \setminus
      T_{i+1}}}$, and suppose that the statement in
  Lemma~\ref{lemma:algksupport} applies, as happens with probability
  $1-e^{\Omega(s)}$.  Then for any level $i$ and $p \in V_i$, if $c_i$
  is the $s$th largest value in $y_{T_i}$, then $y_p \leq
  \max\{f_i/4s, 2c_i\}$ or $y_p \leq \frac{f_i}{4s} + 2c_i$. Since
  $y_{T_i}$ contains at least $s$ values larger than $c_i$, and at
  most $\abs{U \cap T_i} = \abs{U \cap C(T_{i+1})} - \abs{V_i} \leq s
  - \abs{V_i}$ of them lie in $U$, $y_{T_i \setminus U}$ must contain
  at least $\abs{V_i}$ values larger than $c_i$.  This, combined with
  $\abs{V_i} \leq s$, gives
  \begin{align}\label{eq:lostmass}
    \norm{1}{y_{V_i}} \leq f_i/4 + 2\norm{1}{y_{T_i \setminus U}}.
  \end{align}
  Combining Equations~\eqref{eq:subtreesums} and~\eqref{eq:lostmass}, we get
  \begin{align*}
    \norm{1}{y'_{\overline{S}}} &\leq 2[\sum_i \norm{1}{(y-y')_{V_i}} + \norm{1}{y_{V_i}}]\\
    &\leq 2\norm{1}{(y-y')_U} + \sum_{i} \left(4\norm{1}{y_{T_i\setminus U}} + f_i/2\right)\\
    &\leq 2\norm{1}{(y-y')_U} + 4 \norm{1}{y_{S\setminus U}} + \norm{1}{y_{\overline{S}}}/2\\
    &= 2\norm{1}{(y-y')_U} + 4 \norm{1}{(y-y')_{S\setminus U}} + \norm{1}{y_{\overline{S}}}/2\\
    &\leq 4\norm{1}{y-y'} + \norm{1}{y_{\overline{S}}}/2.
  \end{align*}
  Therefore
  \begin{align*}
    \norm{1}{y_{\overline{S}}} &\leq \norm{1}{y-y'} + \norm{1}{y'_{\overline{S}}}\\
    &\leq 5\norm{1}{y-y'} + \norm{1}{y_{\overline{S}}}/2\\
    \norm{1}{y_{\overline{S}}} &\leq 10\norm{1}{y-y'}
  \end{align*}
as desired.
\end{proof}

\subsection{Application to EMD recovery}

By Lemma~\ref{lemma:framework} our $\ell_1/\ell_1$ recovery algorithm
for $\Mc$ gives an $\emd/\emd$ recovery algorithm.

\begin{theorem}\label{thm:randomizedemdemd}
  Suppose $k = \Omega(\log \log n)$.  There is a randomized EMD/EMD
  recovery scheme with $m = O(\frac{1}{\eps^2}k\log(n/k))$, $C =
  \eps$, and success probability $1 - k^{-\Omega(1)}$.  Recovery takes
  $O(\frac{1}{\eps^2}k \log (n/k))$ time.
\end{theorem}

\section{Wavelet-based method}\label{sec:haar}

We can also instantiate the framework of Section~\ref{sec:framework}
using a reweighted Haar wavelet basis instead of $P$ for the embedding
$B$.  We will have $\mathcal{M}$ be the tree-sparse model
$\mathcal{T}_{O(\frac{1}{\eps^2}k \log {n/k})}$, and use the
$\ell_1/\ell_1$ recovery scheme of Section~\ref{sec:model-based}.

\ifacm
\else %
The details are deferred to Appendix~\ref{app:haar}.
We 
\fi %
obtain an
embedding $W$ defined by a Haar transform $H$ (after rescaling the
rows), and the following theorem:

\begin{theorem}
  There exists a matrix $A$ with $O(k \log (n/k))$ rows such that we
  can recover $x^*$ from $Ax$ with
  \[
  \norm{EMD}{x^* - x} \leq C \min_{y \in \mathcal{T}_K}\norm{1}{Wx -
    y} \leq C \min_{k\text{-sparse } x'} \norm{EMD}{x - x'}
  \]
  for some distortion $C = O(\sqrt{\log (n/k)})$.
\end{theorem}

Note that if we ignore the middle term, this gives the same EMD/EMD result as in
Section~\ref{sec:model-based}.  However the middle term may be small for natural  images even if the right term is not.  In particular, it is well known that images tend to be tree-sparse under $H$.

\paragraph{Acknowledgements}
The authors would like to thank Yaron Rachlin from Draper Lab for numerous conversations and the anonymous reviewers for helping clarify the presentation.

{
\bibliographystyle{alpha}
\bibliography{emdsparse}
}

\appendix

\section{Invertibility of Pyramid Transform}\label{app:invertibility}
 If $b$ were $(Px)_S$ for some $S$, then the problem would be fairly
 easy, since $b$ tells us the mass $p_q$ in cells $q$ (in particular,
 if $q$ is at level $i$, $p_q = \frac{b_q}{2^i}$).  Define the
 \emph{surplus} $s_q = p_q - \sum_{r\in C(q)} p_r$ to be the mass
 estimated in the cell that is not found in the cell's children.

We start from the case when all surpluses are non-negative (as is the
case for $(Px)_S$).  In this case, we can minimize $\norm{1}{b-Py}$ by
creating $s_q$ mass anywhere in cell $q$.

\begin{algorithm}
  \caption{Recovering $y$ from $b$ to minimize $\norm{1}{b-Py}$ when
    all surpluses are non-negative.} \label{frompositive} For every cell $q
  \in G$, let $e_q \in \R^n$ denote an elementary unit vector with the
  $1$ located somewhere in $q$ (for example, at the center of $q$).
  Then return
  \[
  y = \sum_{q \in G} s_qe_q.
  \]
\end{algorithm}

\begin{lemma}
  Suppose $b$ is such that $s_q \geq 0$ for all $q \in G$.  Let $y$ be
  the result of running Algorithm~\ref{frompositive} on $b$.  Then $y$
  minimizes $\norm{1}{b-Py}$.
\end{lemma}
\begin{proof}
  The vector $y$ has the property that $(Py)_q \geq b_q$ for all $q \in G$,
  and for the root node $r$ we have $(Py)_r = b_r$.  Because the weights are exponential in the level value, any $y'$ minimizing
  $\norm{1}{b-Py'}$ must have $(Py')_r \geq b_r$, or else increasing any coordinate of $y'$ would decrease $\norm{1}{b-Py'}$.  But then
  \begin{align*}
    \norm{1}{b-Py'} &= \sum_{i=0}^{\log \Delta} \sum_{q \in G_i} \abs{(Py')_q - b_q}\\
    &\geq \sum_{i=0}^{\log \Delta} \sum_{q \in G_i} (Py')_q - b_q\\
    &= \sum_{i=0}^{\log \Delta} \left(2^{i-\log\Delta}(Py')_r - \sum_{q \in G_i} b_q\right)\\
    &= (2 - 2^{-\log \Delta})(Py')_r - \norm{1}{b}\\
    &\geq (2-2^{-\log \Delta})b_r - \norm{1}{b}.
  \end{align*}
  Equality holds if and only if $(Py')_q \geq b_q$ for all $q \in G$
  and $(Py')_r = b_r$.  Since $y$ has these properties, $y$ minimizes
  $\norm{1}{b-Py}$.
\end{proof}

Unfortunately, finding the exact solution is harder when some
surpluses $s_q$ may be negative.  Then in order to minimize
$\norm{1}{b-Py}$ one must do a careful matching up of positive and
negative surpluses.  In order to avoid this complexity, we instead
find a greedy 8-approximation.  We modify $b$ from the top down,
decreasing values of children until all the surpluses are
non-negative.

\begin{algorithm}
  \caption{Modifying $b$ to form all non-negative
    surpluses} \label{makingpositive} Perform a preorder traversal of
  $G$.  At each node $q$ at level $i$, compute the surplus $s_q$.  If
  $s_q$ is negative, arbitrarily decrease $b$ among the children of
  $q$ by a total of $2^{i-1}\abs{s_q}$, so that $b$ remains
  non-negative.
\end{algorithm}

\begin{lemma}
  Suppose we run algorithm~\ref{makingpositive} on a vector $b$ to get
  $b'$.  Then
  \[
  \norm{1}{b - b'} \leq 3 \min_{y} \norm{1}{Py - b}.
  \]
\end{lemma}
\begin{proof}
  Let $y$ minimize $\norm{1}{Py - b}$.  As with $Py'$ for any $y'$,
  $Py$ has zero surplus at every node.

  At the point when we visit a node $q$, we have updated our estimate
  of $b$ at $q$ but not at its children.  Therefore if $q$ is at level
  $i$ we compute $s_q = \frac{1}{2^i}b'_q -
  \frac{1}{2^{i-1}}\sum_{s\in C(q)} b_s$.  Then, because $Py$ has zero
  surplus,
  \begin{align*}
    \abs{s_q} &= \abs{\frac{1}{2^i}b'_q - \frac{1}{2^i}(Py)_q -
      \frac{1}{2^{i-1}}\sum_{s\in C(q)} (b_s - (Py)_s)}\\
    &\leq \frac{1}{2^i}\abs{b'_q - b_q} + \frac{1}{2^i}\abs{b_q - (Py)_q}
    + \frac{1}{2^{i-1}}\sum_{s \in C(q)}\abs{b_s - (Py)_s}.
  \end{align*}
  Define $f_i = \sum_{q \in G_i} \abs{b_q - (Py)_q}$ to be the
  original $\ell_1$ error on level $i$, and $g_i = \sum_{q \in G_i}
  \abs{b'_q - b_q}$ to be a bound on the amount of error we add when
  running the algorithm.  Because we only modify values enough to
  rectify the surplus of their parent, we have
  \begin{align*}
    g_{i-1} &\leq 2^{i-1}\sum_{q \in G_i} \abs{s_q}\\
    &\leq \sum_{q \in G_i} \frac{1}{2}\abs{b'_q - b_q} + \frac{1}{2}\abs{b_q - (Py)_q} + \sum_{s \in C(q)}\abs{b_s - (Py)_s}\\
    &\leq \frac{1}{2}g_{i} + \frac{1}{2}f_{i} + f_{i-1}.
  \end{align*}
  Unrolling the recursion, we get
  \begin{align*}
    g_i &\leq f_i + \sum_{j=1}^{\log \Delta - i} \frac{1}{2^{j-1}} f_{i+j}\\
    \norm{1}{b'-b} = \sum_{i=0}^{\log \Delta} g_i &\leq \sum_{i=0}^{\log \Delta} 3f_i = 3\norm{1}{Py-b}
  \end{align*}
  as desired.
\end{proof}

This lets us prove Lemma~\ref{lemma:pyramidinverse}.

\newenvironment{theorem1}[2]{
    \parskip 0pt 
    \trivlist
    \item[%
        \hskip 10pt
        \hskip \labelsep
        {{\sc #1}\hskip 5pt\ref{#2}.}%
    ]
    \it
}


\begin{theorem1}{Lemma}{lemma:pyramidinverse}
  Given any approximation $b$ to $Px$, running the previous two
  algorithms gives a $y$ with
  \[
  \norm{1}{Py-Px} \leq 8\norm{1}{b-Px}
  \]
  in $O(\abs{\supp(b)})$ time.
\end{theorem1}
\begin{proof}
  By running Algorithm~\ref{makingpositive} on $b$, we get $b'$ with
  $\norm{1}{b - b'} \leq 3\norm{1}{Px - b}$.  Then we run
  Algorithm~\ref{frompositive} on $b'$ to get $y$ that minimizes
  $\norm{1}{Py - b'}$.  Then
  \begin{align*}
    \norm{1}{Py-Px} &\leq \norm{1}{Py-b'} + \norm{1}{Px-b'}\\
    &\leq 2\norm{1}{Px-b'}\\
    &\leq 2(\norm{1}{Px-b} + \norm{1}{b'-b})\\
    &\leq 8\norm{1}{Px-b}.
  \end{align*}
  To bound the recovery time, note that after
  Algorithm~\ref{makingpositive} visits a node with value $0$, it sets
  the value of every descendant of that node to $0$.  So it can prune
  its descent when it first leaves $\supp(b)$, and run in
  $O(\abs{\supp(b)})$ time.  Furthermore, this means $\abs{\supp(b')}
  \leq \abs{\supp(b)}$ and $\supp(b')$ is a top-down tree.  Hence
  Algorithm~\ref{frompositive} can iterate through the support of $b'$
  in linear time.
\end{proof}

\section{Model-based compressive sensing}\label{app:model-based}
In this section we first provide a quick review of model-based sparse
recovery, including the relevant definitions, algorithms and their
guarantees.  We then show how to augment the algorithm so that it
provides the guarantees that are needed for our EMD algorithms.

\subsection{Background}

\paragraph{Model-based RIP}
Given a signal model $\Mc_K$, we can formulate the $\Mc_K$-restricted
isometry property ($\Mc_K$-RIP) of an $m \times n$ matrix $A$, which suffices
for performing sparse recovery.

\begin{definition}
A matrix $A$ satisfies the $\Mc_K$-RIP with constant $\delta$ if for
any $x \in \Mc_K$, we have
\[ (1-\delta)\norm{2}{x}  \leq \norm{2}{Ax} \leq (1+\delta) \norm{2}{x} \]
\end{definition}

It is known that random Gaussian matrices with $m=O(k \log(n/k))$ rows
satisfy the $\Sigma_k$-RIP (i.e., the ``standard'' RIP), with very
high probability, and that this bound cannot be
improved~\cite{DIPW}. In contrast, it has been shown that in order to
satisfy the $\Tc_K$-RIP, only $m=O(K)$ rows suffice~\cite{BCDH}.  The
intuitive reason behind this is that the number of rooted trees of
size $K$ is $2^{O(K)}$ while the number of sets of size $k$ is
$\binom{n}{k} = 2^{\Theta(k\log (n/k))}$.

\paragraph{Algorithms}  Given a matrix $A$ that satisfies the $\Mc_K$-RIP,
one can show how to recover an approximation to a signal from its
sketch.  The specific theorem (proven in~\cite{BCDH} and re-stated
below) considers $\ell_2$ recovery of a ``noisy'' sketch $Ax + e$,
where $e$ is an arbitrary ``noise'' vector, while $x \in \Mc_K$.  In
the next section we will use this theorem to derive an $\ell_1$ result
for a different scenario, where $x$ is an {\em arbitrary} vector, and
we are given its {\em exact} sketch $Ax$.

\begin{theorem}
Suppose that a matrix $A$ satisfies $\Mc^{(4)}_K$-RIP with constant
$\delta<0.1$.  Moreover, assume that we are given a procedure that,
given $y \in \R^n$, finds $y^* \in \Mc_K$ that minimizes
$\|y-y^*\|_2$.  Then there is an algorithm that, for any $x \in
\Mc_K$, given $Ax+e$, $e \neq 0$, finds $x^* \in \Mc_K$ such that
\[ \|x-x^*\|_2 \le C \|e\|_2 \]
for some absolute constant $C>1$. The algorithm runs in time $O(
(n+T+MM) \log(\|x\|_2/\|e\|_2) )$, where $T$ is the running time of
the minimizer procedure, and $MM$ is the time needed to perform the
multiplication of a vector by the matrix $A$.
\label{t:cosamp}
\end{theorem}

Note that the algorithm in the theorem has a somewhat unexpected
property: if the sketch is nearly exact, i.e., $e \approx 0$, then the
running time of the algorithm becomes unbounded.  The reason for this
phenomenon is that the algorithm iterates to drive the error down to
$\norm{2}{e}$, which takes longer when $e$ is small. However, as long
as the entries of the signals $x,x^*$ and the matrix $A$ have bounded
precision, e.g., are integers in the range $1, \dotsc, L$, one can
observe that $O(\log L)$ iterations suffice.

The task of minimizing $\|y-y^*\|_2$ over $y^* \in \Mc_K$ can typically be accomplished in time polynomial in $K$ and $n$.
In particular, for $\Mc_K=\Tc_K$, there is a simple dynamic programming algorithm solving this problem in time $O(k^2n)$.
See, e.g.,~\cite{CIHB} for a streamlined description of the algorithms for (a somewhat more general) problem and references.
For more mathematical treatment of tree approximations, see~\cite{CDDD}.

The following lemma (from~\cite{NT08}) will help us bound the value of $\|e\|_2$.
\begin{lemma}
Assume that the matrix $A$ satisfies the (standard) $\Sigma_{s}$-RIP  with constant $\delta$.
 Then for any vector $z$, we have $\|Az\|_2 \le \sqrt{1+\delta}( \|z_S\|_2 +
\|z\|_1/\sqrt{s})$, where $S$ is the set of the $s$ largest (in
magnitude) coefficients of $z$.
\label{l:l2l1}
\end{lemma}

For completeness, we also include a proof. It is different, and
somewhat simpler than the original one. Moreover, we will re-use one of
the arguments later.

\begin{proof}
We partition the coordinates of $S$ into sets $S_0, S_1, $ $S_2, \dotsc,
S_t$, such that (i) the coordinates in the set $S_j$ are no larger (in
magnitude) than the coordinates in the set $S_{j-1}$, $j \ge 1$, and
(ii) all sets but $S_t$ have size $s$.  We have
\begin{eqnarray*}
 \|Az\|_2 & \le & \sum_{j=0}^t \|A z_{S_j}\|_2\\
 & \le & \sqrt{1+\delta} ( \|z_{S_0}\|_2 + \sum_{j=1}^t \|z_{S_j}\|_2) \\
 & \le & \sqrt{1+\delta} ( \|z_{S_0}\|_2 + \sum_{j=1}^s  \sqrt{s} (\|z_{S_{j-1}}\|_1/s))\\
 & \le & \sqrt{1+\delta} ( \|z\|_2 + \|z\|_1/\sqrt{s})
\end{eqnarray*}
\end{proof}

\subsection{New result}

We start from the following observation relating general sparsity and
tree sparsity.  Consider $k$ and $K$ such that $K=c'k \log(n/k)$ for
some constant $c'$.

\begin{claim}
Assume $n= \frac{c^l-1}{c-1}$ for some (constant) integer $c$. Then
there exists a constant $c'$ such that $\Sigma_k \subset \Tc_K$.
\label{c:st}
\end{claim}

\begin{proof}
It suffices to show that for any $S \subset [n]$ of size
$k$ there exists a rooted connected subset $T$ of $T(c,l)$ of size
$K$ such that $S \subset T$.  The set $T$ is equal to $T' \cup T''$,
where (i) $T'$ consist of all nodes in the tree $T(c,l)$ up to level
$\lceil \log_c k \rceil$ and (ii) $T''$ consists of all paths from the
root to node $i$, for $i \in S$.  Note that $|T'| = O(k)$, and
$|T''\setminus T'| = O(k (\log n - \log k)) = O(k \log (n/k))$.
\end{proof}

This claim is used in the following way.  As we will see later, in order to
provide the guarantee for recovery with respect to the model $\Tc_K$,
we will need to perform the recovery with respect to the model $\Tc_K
\oplus \Sigma_k$.  From the claim it follows that we can instead
perform the recovery with respect to the model $\Tc^{(2)}_K \subset
\Tc_{2K}$.

Specifically, we show the following.

\begin{theorem}
Suppose that we are given a matrix and minimizer subroutine as in
Theorem~\ref{t:cosamp} for $\Tc_{2K}$.  Then, for any $x$, given the
vector $Ax$, the approximation $x^*$ computed by the algorithm in
Theorem~\ref{t:cosamp} satisfies
\[ \|x - x^*\|_1 \le (1 + 2C \sqrt{(1+\delta)c' \log(n/k)})  \min_{x' \in \Tc_K} \|x-x'\|_1
\]
\end{theorem}

\begin{proof}
Let $x' \in \Tc_{K}$ be the minimizer of $\|x-x'\|_1$.  Let $T$ be a
tree of size $K$ such that $x'=x_T$, and define the ``$\ell_1$
approximation error'' $E=\|x-x'\|_1 = \|x_{\overline{T}}\|_1$

 Let $P \subseteq \overline{T}$ be the set of the $k$ largest (in magnitude)
 coordinates of $x_{\overline{T}}$.  By Claim~\ref{c:st} it follows that $P
 \subseteq T'$, for some $T' \in \Tc_{K}$.  Let $T'' = T \cup T'$.


We decompose $Ax$ into $Ax_{T''}+Ax_{\overline{T''}} = Ax_{T''} +e$.
Since $x_{T''} \in \Tc_{2K}$, by Theorem~\ref{t:cosamp} we have
\begin{eqnarray}
\|x_{T''}-x^*\|_2 & \le & C  \|e\|_2
\label{e:xse}
\end{eqnarray}
Let $T^*$ be the support of $x^*$. Note that $|T^*| \le 2K$.

Since $A$ satisfies the (standard) RIP of order $k$ with constant
$\delta=0.1$, by Lemma~\ref{l:l2l1} we have
\[
\norm{2}{e} \le \sqrt{1+\delta}[ \norm{2}{x_S} + \norm{1}{x_{\overline{T \cup P}}}/\sqrt{k}]
\]
where $S \subset \overline{T \cup P}$ is the set of the $k$ largest
(in magnitude) coordinates of $x_{\overline{T \cup P}}$.  By the
definition of $P$, every coordinate of $\abs{x_S}$ is not greater than
the smallest coordinate of $|x_P|$.  By the same argument as in the
proof of Lemma~\ref{l:l2l1} it follows that $\norm{2}{x_S} \le
\norm{1}{x_P}/\sqrt{k}$, so
\begin{align}
\norm{2}{e} &\le \sqrt{(1+\delta)/k}\norm{1}{x_{\overline{T}}}. \label{e:switch}
\end{align}

We have
\begin{align*}
\norm{1}{x-x^*}
& =  \norm{1}{(x-x^*)_{T'' \cup T^*}} + \norm{1}{(x-x^*)_{\overline{T'' \cup T^*}}} \\
& \le  \norm{1}{x_{T''}-x^*} + \norm{1}{x_{T^*\setminus T''}} + \norm{1}{(x-x^*)_{\overline{T'' \cup T^*}}} \\
& =  \norm{1}{x_{T''}-x^*} + \norm{1}{x_{\overline{T''}}} \\
& \le    \norm{1}{x_{T''}-x^*}  + E\\
& \le  \sqrt{4K}   \norm{2}{x_{T''}-x^*}  + E \\
& \le   \sqrt{4K}  C \norm{2}{e} + E \\
& \le    \sqrt{4K}  C \sqrt{(1+\delta)/k} \norm{1}{x_{\overline{T}}}+ E \\
& =    (1 + 2C \sqrt{(1+\delta)K/k})E\\
& =    (1 + 2C \sqrt{(1+\delta)c'\log (n/k)})E
\end{align*}
by Equations~\ref{e:xse} and~\ref{e:switch}.
\end{proof}

\section{Strict sparse approximation}
\label{s:strict}
In this section we show how to reduce the sparsity of an approximation down to $k$ for an arbitrary norm $\|\cdot\|$.
This reduction seems folklore, but we could not find an appropriate reference, so we include it for completeness.

Consider a sparse approximation scheme that, given $Ax$, returns (not necessarily sparse) vector $x^*$ such that
$\|x^*-x\| \le C \min_{k\text{-sparse } x'}  \|x'-x\|$; 
let $x'$ be the the minimizer of the latter expression.
Let $\hat{x}$ be the approximately best {\em k-sparse} approximation to $x^*$, i.e., such that $\|\hat{x} - x^*\| \le C' \min_{k\text{-sparse } x'' } \|x''-x^*\|$; let $x''$ be the minimizer of the latter expression. Note that since $x'$ is $k$-sparse, it follows that $\|x''-x^*\| \le \|x'-x^*\|$.

\begin{claim} We have 
\[ \|\hat{x}-x\| \le  [(C'+1)C+C'] \|x'-x\| \]
\end{claim}
\begin{proof}
\begin{eqnarray*}
\|\hat{x}-x\| 
& \le & \|\hat{x}-x^*\| + \|x^*-x\| \\
& \le & C' \|x''-x^*\| + \|x^*-x\|  \\
& \le & C' \|x'-x^*\| + \|x^*-x\| \\
& \le & C'[ \|x'-x\|+\|x-x^*\| ] + \|x^*-x\| \\
& = & (C'+1)\|x^*-x\| + C'  \|x'-x\| \\
& \le & (C'+1)C\|x'-x\| + C'  \|x'-x\| \\
& = & [(C'+1)C+C'] \|x'-x\|
\end{eqnarray*}

\ifacm
\else
\section{Wavelet-based method}\label{app:haar}

We start by recalling the definition of the {\em non-standard
two-dimensional Haar wavelet basis} (see~\cite{SDS} for an overview).  Let
$H \in \R^{n \times n}$ be the matrix with rows corresponding to the
basis vectors.  We will define $H$ in terms of the grids $G_i$.  The
first row of $H$ has all coordinates equal to $1/n$.  The rest of $H$
consists of three rows for each cell $C \in G_i$ for $i \geq
1$.  For each cell $C$, the corresponding rows contain zeros outside
of the coordinates corresponding to $C$.  The entries corresponding to
$C$ are defined as follows: (i) one row has entries equal to $2^{-i}$
for each entry corresponding to the left half of $C$ and equal to
$-2^{-i}$ for each entry corresponding to the right half of $C$; (ii)
the second row has entries equal to $2^{-i}$ for the top half of $C$
and to $-2^{-i}$ for the bottom half; (ii) and the third row has
entries equal to $2^{-i}$ for the top left and bottom right quadrants
of $C$, and equal to $-2^{-i}$ for the other two quadrants.

We define $W$ to transform into the same basis as $H$, but with
rescaled basis vectors.  In particular, the basis vectors from level
$i$ are smaller by a factor of $2^{2i-2}$, so the non-zero entries
have magnitude $2^{2-3i}$.  This is equivalent to changing the
coefficients of the corresponding rows of $W$ to be $2^{i-2}$ rather
than $2^{-i}$.  Similarly, we rescale the all-positive basis vector to
have coefficients equal to $1/n^3$.  Then $W = DH$ for some diagonal
matrix $D$.

This rescaling is such that the columns of $W^{-1}$, call them $v_i$,
all have $\norm{EMD}{v_i} = 1$.  This is because the min-cost matching
moves each of $2^{2i}/2$ coefficients by $2^{i}/2$.  So we have
\begin{align*}
  \norm{EMD}{x} &= \norm{EMD}{\sum (Wx)_iv_i} \\&\leq \sum \norm{EMD}{(Wx)_iv_i} \notag\\&= \sum \abs{(Wx)_i} = \norm{1}{Wx},
\end{align*}
which is Property~\ref{prop:expansion} of the framework.

Property~\ref{prop:invertibility} is easy since $W$ has a known
inverse (namely $H^T D^{-1}$), giving $\norm{1}{y-WW^{-1}y} = 0$ for
all $y$.  All that remains to show is
Property~\ref{prop:model-alignment}.

\begin{lemma}
  For all $x \in \R_+^{n}$, there exists a $y \in
  \mathcal{T}_{O(\frac{1}{\eps^2}k\log (n/k))}$ with
  \[
  \norm{1}{y-Wx} \leq \eps \min_{k\text{-sparse } x_k}\norm{EMD}{x - x_k}.
  \]
\end{lemma}
\begin{proof}
  We will show this using Lemma~\ref{lemma:pyramid-model} as a black
  box.  We know there exists a support $S$ of $Px$ corresponding to a
  tree of grid cells such that
  \[
  \norm{1}{(Px)_{\overline{S}}} \leq \eps \min_{k\text{-sparse } x_k}\norm{EMD}{x - x_k}.
  \]
  Let $S'$ be a support of $Wx$ that contains the all-constant basis
  vector as well as, for each cell $C \in G_i$ in $S$ with $i \geq
  1$, the three coefficients in $Wx$ corresponding to $C$.  Then $S'$
  is also a tree.

  For any cell $C \in G_i$, let $u$ be the row in $P$
  corresponding to $C$ and $v$ be any of the three rows in $W$
  corresponding to $C$.  Then
  \[
  \norm{\infty}{v} = 2^{i-2} = \frac{1}{4}\norm{\infty}{u}.
  \]
  So the only difference between $v$ and $u$ is that (i) $v$ has one fourth
  the magnitude in each coefficient and (ii) some coefficients of $v$
  are negative, while all of $u$ are positive.  Hence for positive
  $x$, $\abs{v\cdot x} \leq \frac{1}{4}\abs{u\cdot x}$.  This gives
  \[
  \norm{1}{(Wx)_{\overline{S}'}} \leq \frac{3}{4}\norm{2}{(Px)_{\overline{S}}}\leq \frac{3}{4}\eps \min_{k\text{-sparse } x_k}\norm{EMD}{x - x_k}.
  \]
  as desired.
\end{proof}

\begin{theorem}
  This gives
  \[
  \norm{EMD}{x^* - x} \leq C \min_{y \in \mathcal{T}_K}\norm{1}{Wx -
    y} \leq C \min_{k\text{-sparse } x'} \norm{EMD}{x - x'}
  \]
  for some distortion $C = O(\sqrt{\log (n/k)})$.
\end{theorem}

\fi

\end{proof}
\end{document}